# Liquid phase mass production of air-stable black phosphorus/phospholipids nanocomposite with ultralow tunneling barrier

Qiankun Zhang[#,1], Yinan Liu[#,2], Jiawei Lai[2], Shaomian Qi[2], Chunhua An[1], Yao Lu[1], Xuexin Duan[1], Wei Pang[1], Daihua Zhang[1], Dong Sun[2,3], Jianhao Chen[*,2,3] and Jing Liu[*,1]

[1]State Key Laboratory of Precision Measurement Technology and Instruments, School of Precision Instruments and Opto-electronics Engineering, Tianjin University, NO. 92 Weijin Road, Tianjin, China, 300072

[2]International Center for Quantum Materials, School of Physics, Peking University, NO. 5 Yiheyuan Road, Beijing, China, 100871

[3]Collaborative Innovation Center of Quantum Matter, Beijing 100871, China

*Corresponding authors:
Jian-Hao Chen (chenjianhao@pku.edu.cn) and Jing Liu (jingliu-1112@tju.edu.cn)

# These authors contributed equally to this work

# ABSTRACT

Few-layer black phosphorus (FLBP), a recently discovered two-dimensional semiconductor, has attracted substantial attention in the scientific and technical communities due to its great potential in electronic and optoelectronic applications. However, reactivity of FLBP flakes with ambient species limits its direct applications. Among various methods to passivate FLBP in ambient environment, nanocomposites mixing FLBP flakes with stable matrix may be one of the most promising approaches for industry applications. Here, we report a simple one-step procedure to mass produce air-stable FLBP/phospholipids nanocomposite in liquid phase. The resultant nanocomposite is found to have ultralow tunneling barrier for charge carriers which can be described by an Efros-Shklovskii variable range hopping mechanism. Devices made from such mass-produced FLBP/phospholipids nanocomposite show highly stable electrical conductivity and opto-electrical response in ambient conditions, indicating its promising applications in both electronic and optoelectronic applications. This method could also be generalized to the mass production of nanocomposites consisting of other air-sensitive two-dimensional materials, such as FeSe, NbSe$_2$, WTe$_2$, etc.

**Introduction**

Liquid-phase processing is a widely used method in industry for material processing which is easily scalable and potentially low cost [1-4]. It enables mass fabrication of thin film and composite devices for numerous applications, including flexible electronics [5-10], energy harvesting [11-13], environmental monitoring [14-16], water purification [17] and catalysts developments [18]. However, strategies to improve the environmental stability of liquid processed materials are seldom studied, which become crucially important due to the emergence of a series of high performing low-dimensional materials with low environmental stability [19-23]. Black phosphorus (BP) is one of such important but air-sensitive materials. It is expected to have great potential in electronic and optoelectronic applications, because of its high carrier mobility [24], tunable direct bandgap [25-27], and electrical/optical/mechanical anisotropy [27-30]. Thus liquid-phase exfoliation/ processing of few-layer BP (FLBP) has attracted considerable attentions [31-32]. High quality exfoliation and suspension has been successfully achieved in both organic and aqueous solvents [3,4,33]. However, it remains a challenge to achieve air stability for the mass production, since currently available passivation techniques [34, 35] are not directly compatible with the liquid phase process. Consequently, the existing mass production techniques are not yet suitable to be used for industrial-scale applications.

In this work, we accomplished FLBP mass exfoliation and passivation in a single step by exfoliating bulk BP in chloroform with phospholipids (1,2-dipalmitoyl-sn-glycero-3-phosphocholine, shortened as DPPC) dissolved in. DPPC is the main component of cell membrane and its molecular structure contains a hydrophilic tail and a hydrophobic head. The amphiphilic nature enables these molecules to intercalate with bulk BP crystals to improve exfoliation efficiency of the later (shown in Fig. 1a), and then self-assemble a thin layer on the surface of exfoliated FLBP to protect it from ambient environment (as shown in Fig. 1b). No substantial morphological changes, as measured by atomic force microscopy (AFM), are observed for these DPPC passivated FLBP kept in air for more than two weeks. More importantly, the electrical resistance of the FLBP/DPPC composite remained almost the same after exposure to ambient conditions for one month (Fig. 2f). We also found that the insulating DPPC [36] passivation produces a surprisingly weak barrier for charge carrier transport. Temperature-dependent measurement reveals an Efros-Shklovskii variable range hopping (ES-VRH) mechanism for the FLBP/DPPC composite which exceeds the most optimistic expectation from the rigid BP/DPPC/DPPC/BP tunneling model.

**Methods**

*Preparation of DPPC exfoliated FLBP.* The DPPC solution was prepared by adding 30 mg DPPC (in powder form, purchased from Sigma-Aldrich Inc., Part No. P0763) into 100 mL chloroform. Then, 30 mg BP powder was added in 100 mL of the prepared DPPC solution and exfoliated in a kitchen blender for 30 min. Then, the exfoliated suspension was centrifuged for 30 min at 1000 rpm. The supernatant was formed on the top of the suspension. Top 80% of the supernatant containing DPPC wrapped FLBP was collected and lyophilized for further analysis. All the above procedures were conducted under ambient conditions. The concentration of FLBP was estimated to be 0.28 mg/mL by measuring the mass of the lyophilized powder of the top 80% supernatant after centrifugation.

*Sample characterization.* HR-TEM and EELS images were taken with a Tecnai G2 F20 (FEI, USA). HR-TEM and EELS samples were prepared by pipetting several microliters of DPPC/FLBP dispersion onto the holey carbon grid. Raman spectra were acquired by DXR microscopy (Thermo Scientific, USA) with a 532 nm laser. The sample was prepared by drop-casting FLBP/DPPC onto Si/SiO$_2$ substrate. AFM measurements were performed using a Dimension Icon (Bruker, German) instrument operating in tapping mode. A mechanically exfoliated FLBP was transferred onto Si/SiO$_2$ substrate followed by dip-coating a layer of DPPC on its surface. All the sample characterizations were performed under ambient conditions.

*FLBP/DPPC composite device fabrication.* Interdigitated electrodes were pre-fabricated on lithium phosphate (300 μm) substrate for transport and photocurrent measurements (photocurrent data is shown in supplementary information), by standard UV lithography and lift-off processes. The electrodes were made of 20 nm Cr and 180 nm Au. The width and length of each electrode finger were 5 μm and 500 μm, respectively, with a lateral spacing of 15 μm. Then, FLBP/DPPC dispersion was drop casted onto the interdigitated electrode by a micropipette. After the solvent was dried in ambient condition, a thin film of FLBP/DPPC composite was formed on the top of electrodes.

*Temperature dependent resistance measurement.* Transport measurements were performed with a probe station in ambient conditions and with a Quantum Design Physical Property Measurement System (PPMS) at low temperatures. Electrical excitation was applied by a Keithley 2400 source meter controlled by a Labview program written in-house. No gate voltage was applied during transport measurements.

**Results and Discussion**

**Sample characterization.** FLBP was exfoliated in chloroform with and without DPPC by a kitchen blender. The blender was accelerated up to 24000 revolutions per minute (rpm) to generate a fully developed turbulent flow within the container, which provides much higher exfoliation efficiency than standard sonication method [2,4]. Centrifugation was then performed at 1000 rpm for 30 min to disperse FLBP uniformly. Top 80% of the supernatant was collected and used for further testing. Figure 1c shows the exfoliation results of 30 mg BP dissolved in 0.3 mg/mL DPPC chloroform solution (right) and pure chloroform (left), respectively, under the same exfoliation power and duration. The one exfoliated in 0.3 mg/mL DPPC chloroform solution showed much darker color, and thus had much higher concentration of FLBP in the dispersion, compared to the one exfoliated in pure chloroform. The concentration of FLBP/DPPC dispersion was measured to be around 0.28mg/ml, which is the same order of magnitude as the concentration exfoliated by other organic solvent [2,3,4].

Furthermore, several standard techniques were applied to characterize the DPPC exfoliated FLBP flakes. The successful exfoliation of BP in DPPC chloroform solution was confirmed by high resolution transmission electron microscopy (HR-TEM), as shown in Fig.1d and 1e. The TEM image shows that the lattice of the crystal is intact over a 400-nm by 400-nm region, indicating that high quality FLBP flakes can be obtained using DPPC assisted liquid-phase exfoliation. Raman spectroscopy was used to further confirm the crystal structural integrity of the exfoliated FLBP nanosheets (Fig. 1h). The

Raman spectrum presents the characteristic $A_g^1$, $B_{2g}$ and $A_g^2$ phonon vibration modes at wavenumbers of ~359, ~436, and ~463 cm$^{-1}$, respectively.. These peak positions are consistent with the Raman spectra of the mechanically exfoliated FLBP flakes [2,3,4], suggesting that the flakes are crystalline after the liquid phase exfoliation. In order to verify the coverage of FLBP with DPPC, electron energy loss spectroscopy (EELS) was performed to characterize the DPPC exfoliated FLBP nanosheets. Figure 1f and g are the EELS images of a DPPC exfoliated FLBP flake presented in the inset of Fig. 1f. The images show the spatial distribution of phosphorus and nitrogen element, respectively. The EELS micrographs present the homogeneous distribution of both phosphorus and nitrogen elements (from DPPC), confirming that DPPC uniformly covers the FLBP.

Figure 1i, j and k exhibit the controlled size and thickness distribution of the exfoliated FLBP by centrifuging the FLBP dispersion under different speeds of revolution. Three DPPC/FLBP dispersion solutions were prepared by performing the centrifugation at revolutions of 6000, 4000 and 2000 rpm, respectively. After centrifugation, a droplet collected from top 80% of each supernatant was drop casted on Si/SiO$_2$ substrate for further characterization. The diameters and thicknesses of the FLBP flakes were measured by atomic force microscopy (AFM), the distributions of which both followed the Gaussian function. For the samples centrifuged at 6000, 4000 and 2000 rpm, the distributions of the diameters were centered at 1.3 nm, 200 nm and 1100 nm, respectively. The thickness distributions were centered at 4 nm and 9 nm for samples centrifuged at 4000 rpm and 2000 rpm, respectively. This capability to control the size and thickness distributions is vital to applications such as advanced batteries, sensors, light emitting diodes, biological therapy products, when FLBPs with suitable size ranges are required.

The passivation effect of DPPC on FLBP was examined by investigating the stability of morphology and electrical resistance of FLBP/DPPC under long-term air exposure. The morphology was measured by AFM on a mechanically exfoliated FLBP flake transferred onto a Si/SiO$_2$ substrate which was then covered by a DPPC layer through a dip-coating process (Since additional DPPC molecules randomly spread on top of the self-assembly layer, the surface of FLBP flakes obtained directly by liquid exfoliation technique is very rough and hence hard to determine the effect of DPPC on FLBP air stability by morphology). Figure 2a-c show the AFM images of this flake exposed in air for 0, 7 and 16 days, respectively. The flake does not show obvious sign of oxidation, in stark contrast to the bare FLBP flake (Fig. 2d and e). We selected a square area of 1 μm × 1μm in each AFM image as marked by white broken lines in Fig. 2a-e and measured the surface roughness of each of them. The surface roughness of DPPC passivated FLBP flake increases slightly after more than two-week air exposure [37], while the surface roughness of non-passivated FLBP exhibits an eight-fold increase in just 3 days [38], as presented by red lines in Fig. 2f. To study the effect of DPPC on the electrical resistance of FLBP/DPPC under ambient conditions, the device was prepared by drop-casting the liquid-phase exfoliated FLBP/DPPC dispersion on the interdigitated electrode pre-fabricated on lithium phosphate substrate (as shown in Fig. 2f). Remarkably, the resistance of the composite remains almost constant (only ~7% changes) for twenty-eight days, with a value around 100kΩ, except it decreased by about 25% for the first two days [39]. The change within the first two days was likely due to the evaporation of organic solvent. In contrast, the resistance of the device fabricated by bare FLBP dispersion in the same manner never decreases, but rather, increases almost nine orders of magnitude within 5 days (Fig. 2f). Both the morphology analysis and electrical resistance measurement demonstrate the capability of DPPC layer to isolate BP from water and/or oxygen in the air.

Besides the excellent passivation capability, another interesting property of the DPPC layer was observed after careful examination of the electrical conductivity of the FLBP/DPPC nanocomposite: the DPPC layer does not electrically insulate the encapsulated FLBP. Typically, for electrical transport between two conductors separated by a single rigid insulating layer (barrier), either an increase in the barrier thickness at constant bias or a decrease in the bias voltage at constant barrier thickness leads to an exponential increase in the resistance. If the number of barriers increases in a device, under constant bias voltage the effective bias across each barrier will decrease. Based on this model, devices made from the FLBP/DPPC composite should be completely insulated, given the thickness of the DPPC molecule that define the barrier thickness and the average size of the exfoliated BP flakes that determine the number of barriers in the device (see supplementary information for details). However, the resistance of the

FLBP/DPPC composite is only 100 kΩ and remains conductive down to 100 K as discussed below in the *Transport mechanisms*. This result indicates that the effective barrier for charge carriers to tunnel through the DPPC layer is low. Thus, the DPPC layer cannot only drastically increase the stability of exfoliated FLBP in ambient environment, but also preserve its electrical conductivity, which is crucial for electronics applications based on mass production of FLBP.

As the main component of cell membrane, phospholipid layer maintains the integrity of cells and prevents exotic molecules from entering into the cell because of their amphiphilic nature and self-assembly properties. Therefore, it is not surprising that DPPC can cover the FLBP surface uniformly, and prevent water and/or oxygen molecules from reaching BP surface. Additionally, as the interconnection among the phospholipid molecules is very weak, the phospholipid layer can keep the cell fluidity. As a result, the DPPC layer may not be a rigid insulating layer, instead, it allows carriers easily transport between two BP flakes and from BP flakes to the electrodes. This could explain the exceedingly low resistance of the FLBP/DPPC composite in contrast to the estimation from the rigid tunneling barrier model (more details in supplementary information).

**Transport mechanism.** We carried out temperature dependent resistance measurement of the FLBP/DPPC composite to determine its carrier transport mechanism. Figure 3a shows the $I_D$-$V_D$ curves of the device fabricated by depositing FLBP/DPPC dispersion on interdigitated electrode which was measured at temperatures $T$ ranging from 100 K to 200 K. The two-probe resistance of the FLBP/DPPC composite is around 10 GΩ below 100 K, which decreases rapidly as temperature increases.

In order to reveal the nature of charge transport in such composite, we investigated $I_D$ vs. $T$ curves at several $V_D$ ranging from 2V to 7V (see Fig. 3b). It shows that $I_D$ exhibits an exponential activation behavior as the temperature increases. Based on the fact that the device consists of multiple FLBP flakes overlapping each other, the charge carriers will need to hop from one FLBP flake to another during the transport process. Thus we adopt the variable range hopping (VRH) conduction model which generally describes charge carrier conduction in disordered systems:

$$R(T) = R_0 \exp\left(\frac{T_0}{T}\right)^p \tag{1}$$

where $R_0$ is the pre-factor, $T_0$ and $p$ is the characteristic temperature and exponent, respectively. *p* is the important parameter whose value helps to distinguish different charge carrier hopping mechanisms, as described below. In the more ordinary case of the Mott-VRH model, the density of states near the Fermi surface is taken as a constant and the value of $p$ in Eq. (1) is given by $p=1/(D+1)$, where $D$ is the dimension of the system [40-42]. In a two dimensional (2D) system, $p=1/3$, and the characteristic temperature $T_0$ is:

$$T_0 \equiv T_M = \left(\frac{3}{k_B N(E_F)\zeta^2}\right) \tag{2}$$

where $N(E_F)$ is the density of states near the Fermi energy $E_F$, and $\zeta$ is the localization length. In the case of strong Coulomb [43, 44] interactions or in systems with multiple overlapping semiconductor nanoflakes [45, 46], the behavior of the charge transport can be described by the Efros-Shklovskii (ES) VRH model. In such case, the density of states vanishes linearly with charge carrier energy approaching $E_F$ and a Coulomb gap appears at the Fermi level. The exponent $p$ in the ES-VRH is 1/2 in all dimensions [47, 48]. The characteristic temperature in 2D is then given by:

$$T_0 \equiv T_{ES} = \left(\frac{2.8 e^2}{4\pi\varepsilon\varepsilon_0 k_B \zeta}\right) \tag{3}$$

where $\varepsilon_0$ and $\varepsilon$ are the permittivity of vacuum and relative dielectric constant of the material, respectively.

In lightly doped semiconductors, the Coulomb gap is the same order of magnitude with the measured temperature range. When temperature becomes higher, thermal effects will smear out the Coulomb gap and a crossover from ES- to Mott-VRH with increasing temperature is often observed. On the other hand, for $T < T_{ES}/(2.8^2\pi)$, the Coulomb interaction plays a dominant role, and only ES's law is expected [49].

Here, we first determine the key exponent $p$. From Eq. (1), one can obtain the reduced activation energy $W$ [50, 51]:

$$W = -\frac{\partial \ln R(T)}{\partial \ln T} = p \times \left(\frac{T_0}{T}\right)^p \tag{4}$$

The value of $p$ can be obtained from $\ln W = A - p \ln T$, where $A$ is a constant. Figure 4a, b and c show $\ln W$ vs. $\ln T$ plots at bias voltage $V_D = $ 3V, 4V and 5V, respectively. From the linear fits, the value of $p$ is found to be 0.471, 0.454, and 0.484 for $V_D = $ 3, 4 and 5V, respectively. These $p$ values match very well with the ES-VRH model. Hence in the following discussion, we use the ES-VRH model to analyze the temperature dependent resistance data.

Figure 4d shows a semi-log scale plot of $R$ vs. $T^{-1/2}$ for the device at different bias voltages. The linearity of the log resistance with inverse square root of temperature for all the bias voltages in Fig. 4d shows that $R(T)$ fits very well with the $T^{-1/2}$ behavior. By linearly extrapolating the solid lines in Fig. 4d to high temperature limit, $R_0$ for all bias voltages can be obtained and the range of $R_0$ is from 320.5 to 78.0 $\Omega$.

From the slopes of Fig. 4d, the characteristic temperature $T_{ES}$ for the device at different bias voltages can be obtained. Using an effective relative dielectric constant $\varepsilon = $ 8.3 for BP flakes [52-54], the value of $T_{ES}$ is determined to be approximately $3 \times 10^4$ K for the bias voltages ranging from 2V to 7V in one volt interval. $T_{ES}$ is much higher than room temperature, which satisfies the criterion $T < T_{ES}/(2.8^2 \pi)$ [49], and is consistent with the fact that Mott-VRH is not observed in our devices [50, 51]. From the values of $T_{ES}$ and using Eq. (3), the localization length ($\zeta$) is determined to be between 0.17 nm to 0.19 nm, which is a description of the tunneling process between the BP flakes and does not depend on the size of the flakes [52]. A summary of the results obtained from our measurements is presented in Table 1.

**Conclusions**

In conclusion, we accomplished single-step mass production of FLBP/DPPC nanocomposite in liquid phase. The self-assembled DPPC layer on the surface of FLBP during liquid exfoliation prevents water and/or oxygen from reaching FLBP, and thus greatly improves the air-stability of the passivated FLBP. On the other hand, DPPC exhibits a surprisingly low tunnel barrier for charge transport, resulting in relatively low resistance of the FLBP/DPPC composite. Temperature dependent resistance measurement of the FLBP/DPPC composite reveals an Efros-Shklovskii variable range hopping mechanism. We anticipate the investigations and characterizations performed on FLBP/DPPC composite in this work help to promote the mass production of thin film devices based on BP and other air-sensitive nanomaterials.

**Acknowledgement**

This project has been supported by the National Basic Research Program of China (973 Grant Nos. 2013CB921900, 2014CB920900), the National Natural Science Foundation of China (NSFC Grant Nos. 21405109, Nos. 11374021 and Nos. 11674013), and Seed Foundation of State Key Laboratory of Precision Measurement Technology and Instruments (Pilt No. 1710).

**Author Contributions**

J.L. and J.C. conceived the experiment with help of D.S. Q.Z. exfoliated the BP thin flakes, completed TEM, EELS, Raman and AFM measurements, fabricated and tested the devices under the supervision of J.L.. C.A., Y.L. performed dip-coating process. X.D., W.P. and D.Z. provided experimental instruments for dip-coating and device fabrication and testing. Y.L. performed the transport measurements under the supervision of J.C. J.L. performed the photocurrent measurements under the supervision of D.S. Y.L., Q.Z., D.S., J.L and J.C. discussed the results, analyzed the data and wrote the manuscript. All the authors commented on the paper.

99.

**TABLE 1**. Summary of ES-VRH fitting results with various biases.

| $V_D$(V) | 2 | 3 | 4 | 5 | 6 | 7 |
|---|---|---|---|---|---|---|
| $R_0$(Ω) | 320.5 | 178.1 | 137.0 | 102.9 | 84.8 | 78.0 |
| $T_0$(K) | 29433 | 30772 | 30867 | 31400 | 31826 | 31769 |
| $\zeta$(nm) | 0.191 | 0.183 | 0.182 | 0.179 | 0.177 | 0.177 |

# CAPTIONS

**Figure 1. Characterizations of liquid exfoliated BP with DPPC.** a) Schematic diagram of DPPC intercalating bulk BP crystal during liquid exfoliation. b) Schematic diagram of DPPC self-assembly layer on the surface of FLBP; c) Dispersions of FLBP nanoflakes exfoliated in chloroform in the absence (left) and presence (right) of DPPC under the same conditions. d) HR-TEM image of a FLBP exfoliated by DPPC in chloroform (scale bar is 50 nm) and e) FFT pattern of the selected area represented by red box in (d) (Scale bar is 5 nm). f) and g) EELS images of the FLBP flake presenting nitrogen and phosphorus elements, respectively, are obtained from HR-TEM image shown in inset of g (scale bar is 20 nm). h) Raman spectrum of exfoliated FLBP nanoflake, with signature peaks observed at ~359, ~436, and ~463 $cm^{-1}$, respectively. i) Size distribution histogram of BP dispersion centrifuged at 2000 rpm, 4000 rpm, and 6000 rpm, respectively. j) and k) Thickness distribution histogram of BP dispersion centrifuged at 2000 rpm and 4000 rpm, respectively.

**Figure 2. Air stability of DPPC wrapped FLBP.** a), b) and c) AFM images of the same mechanically exfoliated FLBP flake covered by DPPC layer kept in ambient condition for 0 day, 7 days and 16 days, respectively. d) and e) AFM images of mechanically exfoliated bare FLBP flake kept in ambient condition for 0 day and 3 days, respectively. The square areas marked by white broken lines in Fig. 2a-e (with size of 1μm × 1μm) were selected for surface roughness measurement. All scale bars are 1μm. f) Time-dependent evolutions of surface roughness and resistance. Surface roughness of FLBP with and without DPPC passivation was measured in the square areas in Fig. 2a-e (presented by red lines in upper and right axis); resistance measurements consist of bare FLBP and FLBP/DPPC nanocomposite devices (presented by black lines in lower and left axis). The diameter and thickness of the composite (FLBP) film was around 25 (15) μm and 2 (2) μm, respectively. Inset shows the cartoon of the device used for resistance measurements.

**Figure 3. Temperature dependent resistance measurement of the FLBP/DPPC nanocomposite.** a) Current *vs.* voltage curves of the FLBP/DPPC nanocomposite device

in the temperature range of 100-200 K. Inset: Photo of FLBP/DPPC film deposited on interdigitated electrodes. b) Current *vs.* temperature curves for the device at voltage bias ranging from 2V to 7V.

**Figure 4. Efros-Shklovskii Variable ranging hopping of the FLBP/DPPC composite.** (a)–(c) Reduced activation energy ($W$) plotted *vs.* temperature ($T$) in a log-log scale of the device with voltage bias set at 3, 4 and 5V respectively. The squares are the experimental data points and the solid red lines are linear fit of the data. From the slopes of the plots, we obtain $p = 0.454$, $0.448$, and $0.451$ for $V_D = 3$, 4 and 5V, respectively. All correspond to the ES-VRH model. (d) Semi-log scale plot of $R$ *vs.* $T^{-1/2}$ for the device at different voltage biases. From the slopes, we obtain $T_{ES}$ =29433, 30772, 30867, 31400, 31826 and 31769 K for the voltage bias ranging from 2V to 7V, respectively.

**FIGURES**

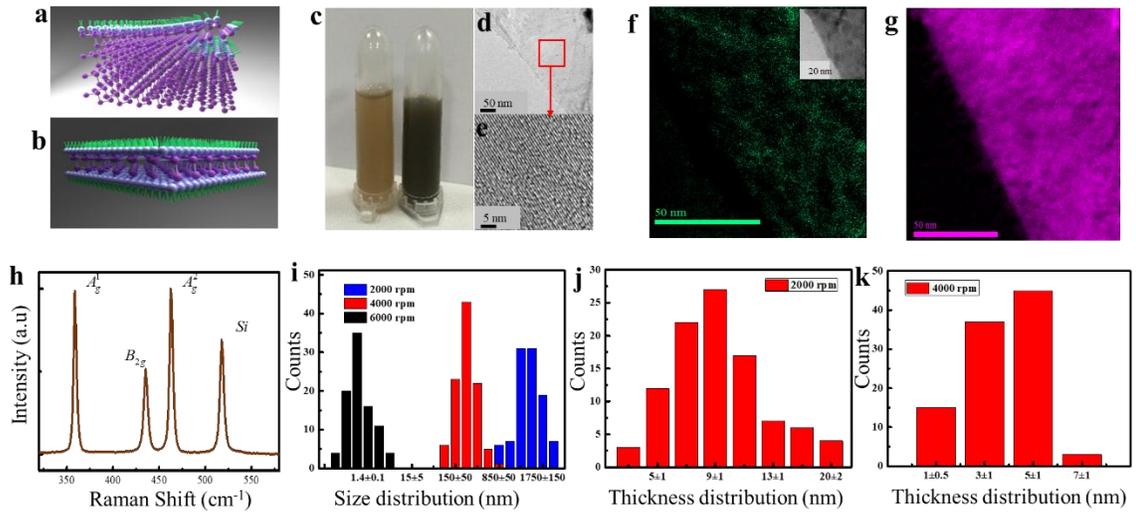

**Figure 1**

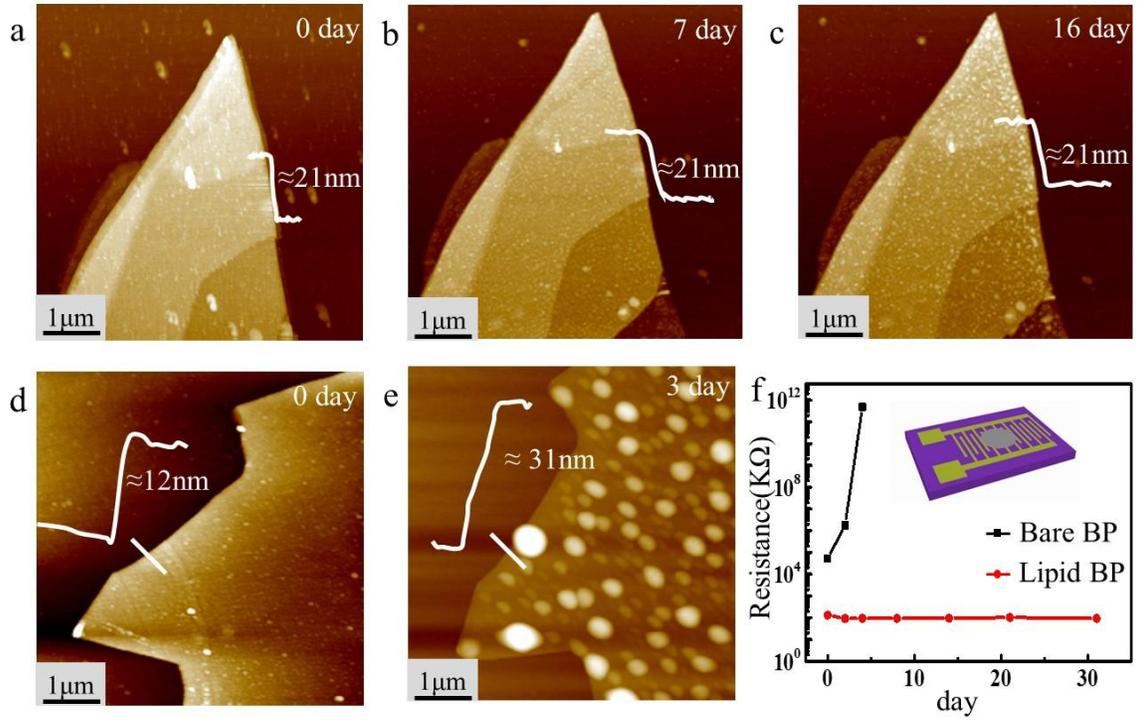

**Figure 2**

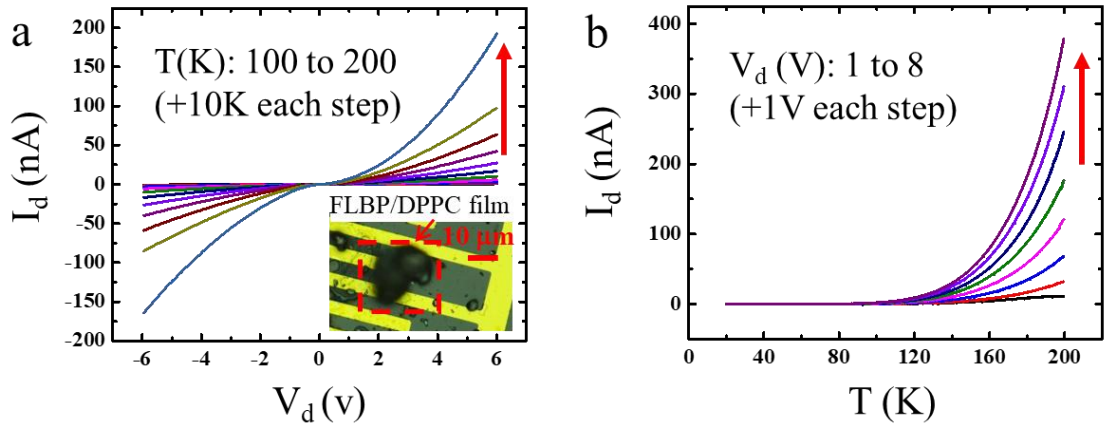

**Figure 3**

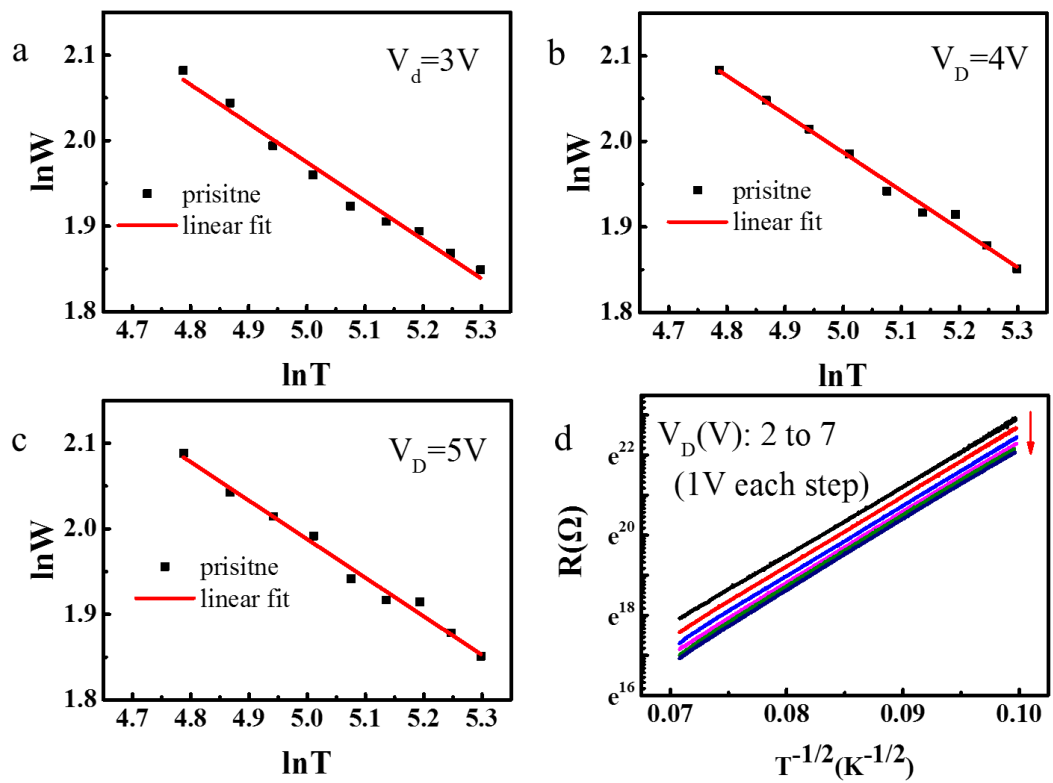

**Figure 4**